\documentclass[review]{elsarticle}
\usepackage{booktabs}
\usepackage{lineno,hyperref}
\usepackage{array,multirow}
\usepackage{graphicx}
\usepackage{amssymb}
\usepackage{tabularx}
\modulolinenumbers[5]
\usepackage{amsmath}
\usepackage{subcaption}
\usepackage{tablefootnote}

\usepackage{dcolumn,booktabs}
\newcolumntype{d}[1]{D..{#1}}
\newcommand\mc[1]{\multicolumn{1}{c}{#1}}
\journal{ Solid State Communications }

\begin{document}

\begin{frontmatter}

\title{Excitation energies and UV-Vis absorption spectra from INDO/s+ML}


\author{Ezekiel Oyeniyi }
\address{Department of Physics, University of Ibadan, Nigeria\\
email: eoyeniyi22@gmail.com, eo.oyeniyi@ui.edu.ng}

\author{Omololu Akin-Ojo }
\address{Department of Physics, University of Ibadan, Nigeria\\
email: oakinojo@gmail.com, o.akin-ojo@ui.edu.ng}


\begin{abstract}
The semi-empirical INDO/s method is popular for studies of excitation energies
and absorption of molecules due to its low computational requirement, making it
possible to make predictions for large systems. However, its accuracy is
generally low, particularly, when compared with the typical accuracy of other
methods such as time-dependent density functional theory (TDDFT). Here, we present machine learning (ML) models that correct the INDO/s results with negligible increases in the amount of computing resources needed.   While INDO/s
excitations energies have an average error of about 1.1 eV relative to TDDFT
energies, the added ML corrections reduce the error to 0.2 eV.  Furthermore, this combination of INDO/s and ML produces UV-Vis absorption spectra that are in good agreement with the TDDFT predictions.
\end{abstract}


\end{frontmatter}


\section{Introduction}
UV/VIS/NIR spectroscopy is an effective tool for studying the optical response properties of materials. It is also useful
in the characterization of compounds. This spectroscopy can detect electromagnetic radiation ranging from 200 to 3000 nm (0.4 to 6.2 eV), primarily caused by electronic excitations. UV/VIS/NIR spectra can be challenging to recognize, interpret, or assign depending on the system under observation. This is particularly true for systems with many ($\gtrapprox 20$)
atoms. This is one place where theoretical and computational studies come
in – to aid in describing and interpreting experimental UV/VIS/NIR spectra. 
However, computational determination of UV-VIS spectra is not without its challenges.
Determination of accurate calculated spectra is computationally intensive, especially for large
systems with many $(\gtrapprox 200)$ electrons. Invariably, one has to use the time-dependent density functional theory (TDDFT)\cite{marques2012fundamentals,casida2009time} method for such big compounds.
The accuracy of TDDFT is on the order of $0.2$~eV which is less than that for GW~\cite{van2012gw}, Quantum Monte Carlo (QMC),  Bethe-Salpeter
equation (BSE) solutions ~\cite{nakanishi1969general,leng2016gw,blase2018bethe}, and the Equation-of-Motion Coupled Cluster with Single and Double excitations (EOM-CCSD) method~\cite{christiansen1995second}. The Complete active-space second-order perturbation theory (CASPT2)~\cite{andersson1990second}, Configuration
interaction with singles (CIS) and doubles [CIS(D)]
excitations~\cite{krishnan1980derivative} have about the same accuracy as TDDFT.
Semi-empirical approaches are even less accurate than CIS, CISD, or TDDFT and
they are used mainly for large systems in order to obtain qualitative or semi-quantitative results for cases in which TDDFT is computationally
prohibitive. For huge
systems, the popular fallback semi-empirical approach for excited states calculations is the INDO/s method~\cite{ridley1973intermediate, inbookINDOs}. INDO/s is used even though it is not as accurate
as the other (more compute intensive) approaches listed above. Increasing the accuracy of the INDO/s method without significantly affecting its computational cost is the goal of this paper. 
This work presents a technique that leads to accuracy similar to that of TDDFT but
at the computational cost of INDO/s. This is achieved by adding machine learning
(ML) derived corrections to INDO/s results.

In the past, we developed a method that systematically re-optimizes the paramaters of the
INDO/s semi-empirical method in order to achieve high
accuracy~\cite{oyeniyi2019efficient, oyeniyi2023oeindo}. However, this re-parameterization has to be done for each unique pair of atoms. While this is possible, a cheaper alternative is to add a correction to the existing semi-empirical methods. The correction we are proposing is based on ML techniques.

Different authors have used ML methods to directly determine
UV-VIS spectra.~\cite{xue2020machine, ghosh2019deep, vinod2023multifidelity,
ramakrishnan2015electronic, westermayr2020machine} The approach has been successful to different
extents. However, since spectra are computed based on quantum mechanics, an approach with a basis in quantum mechanics, upon which ML methods are built or added, should lead to better transferability, i.e., work accurately for many systems for
which it was not trained specifically on. Ramakrishnan {\it et. al.} added
corrections to TDDFT using machine learning in order to reproduce spectra at the
second-order approximate coupled-cluster (CC2) level~\cite{ramakrishnan2015electronic}.  While their method was successful for excitation energies, relatively poorer performance was found for oscillator strengths. Most recently, de Armas-Morejón and co-workers introduced an ML approach to efficiently predict the spectroscopic properties of organic molecules. They proposed a combination of low-cost electronic descriptors derived from DFT calculations—specifically orbital energy differences, Kohn-Sham transition dipole moments, and a so-called charge-transfer character to train ML models. They demonstrated that these electronic descriptors, unlike purely geometrical ones, enable Neural Networks to accurately estimate the absorption spectrum, excitation energy, oscillator strength, and charge-transfer character, achieving results close to high-level TDDFT results at a small fraction of the computational cost~\cite{de2023electronic}.

The next section introduces our Machine Learning correction approach to a semi-empirical method. This is followed by a description of the methods, ML models, descriptors, and data used. We discuss our results in Section~\ref{sec:Results}. Finally, we give a conclusion and recommendation for further studies.

\section{Methods}

\subsection{ML model}

We computed the absorption spectrum, \( \varepsilon(\omega) \), of molecules for electronic transitions at frequency $\omega$ using the expression \cite{barone2011computational}:

\[
\varepsilon(\omega) = \frac{\pi^2 e^2 N
_A}{\ln{(10)}2\pi \epsilon_o n m_e c} \sum_{i,j} f_{ij} g(\omega-\omega_{ij})
\]

where $e$ is the electronic charge, $N_A$ is Avogadro's number, $n$, taken to be unity,  is the refractive index, $m_e$ is the mass of the electron, $\omega$ is the angular frequency and $\omega_{ij} = (\epsilon_i - \epsilon_j)/\hbar$ for transition from state $j$ with energy $\epsilon_j$ to state $i$ with energy $\epsilon_i$. The oscillator strength $f_{ij}$ is given by 

\[
f_{ij} = \frac{2m_e}{3\hbar e^2} \omega_{ij} |\mu_{ij}|^2.
\]
Here, $\mu_{ij}$ is the transition dipole moment for transition from the quantum state $j$ to state $i$. 
The lineshape function, $g(\omega-\omega_{ij})$ could either be a Lorentzian or Gaussian function. Here, we use the Gaussian function:
\[
g(\omega-\omega_{ij}) =  \frac{1}{\sqrt{ 2\pi \delta^2}} e^{ -\frac{(\omega-\omega_{ij})^2}{2\delta^2} } 
\]
where $\delta$ is the broadening factor.
Predicting absorption spectrum accurately depends on precise properties, including excitation energies and oscillator strengths (which are related to the transition dipole moments). In this work, we determined more accurate properties by adding corrections to the properties obtained from an inexpensive low-level theory.

Our property, specifically the excitation energy or oscillator strength, \( P_t \), is estimated using the equation:

\[ P_t = P_b + \Delta_{ML} \]

where \( P_b \) represents the low-level property obtained from INDO/s, and \( \Delta_{ML} \) is the correction, obtained using ML, that accounts for the difference between the high-level and low-level property. The high-level method used was TDDFT and INDO/s for the low level one. 

The ML correction \( \Delta_{ML} \) was developed with various machine learning models namely, Kernel Ridge Regression (KRR) \cite{murphy2012machine}, Support Vector Regression (SVR)\cite{scholkopf2000new}, Multi-Layer Perceptron (MLP), and Random Forest (RF) \cite{breiman2001random}, as implemented in the Scikit-Learn library \cite{scikit-learn}. Detailed reports for the best performing models (KRR and RF) are given here, while the results of the others are provided in the Supplementary Information. 


\subsection{Descriptors}
Descriptors (also known as features) are molecular properties, which are represented in a format suitable for optimal learning, prediction, and inference by machine learning algorithms. In this study, the Coulomb matrix (CM) \cite{rupp2012fast,  hansen2013assessment, collins2018constant} and connectivity counts (Co)\cite{collins2018constant} were used as the features. These descriptors were obtained from atomic coordinates and nuclear charges of molecules, {\it vide infra}. These two features gave the most accurate models of all the ones we tried. 

The Coulomb matrix and its variant, bags of bonds, are among the most commonly used descriptor representations. They have proven to be highly effective but are size-dependent, which limits their application to small molecules.
The Coulomb matrix ($CM$)was one of the two descriptors that gave the best accuracy. Its elements $CM_{ij}$ are defined as follows:~\cite{collins2018constant}

\[
CM_{ij}= 
\begin{cases}
Z_i^{0.5} & \text{if } i=j \\
\frac{Z_i Z_j}{|{\vec r}_i - {\vec r}_j|} & \text{if } i \neq j   
\end{cases}
\]
In this expression, \( Z_k \) and \( {\vec r}_k \) are the atomic number and coordinates, respectively, of the $k$-th atom. 

For the connectivity count (Co) descriptors, feature vectors are created by counting the occurrences of specific bonding patterns.~\cite{collins2018constant} In particular, the so-called Rank-3 feature $Co_{(T_1,T_2)}$ tells us how many $T_1-T_2$ bonds are in the molecule. If a central atom $j$ is bound to atom $i$ by a $T_1$ bound and to atom $k$ by a $T_2$ bound, then this counts as one $T_1-T_2$ bond in the system. Specifically:\cite{collins2018constant} 
\[
Co_{(T_1, T_2)} = \sum_{i<j<k}\delta_{B(i,j),T_1}\delta_{B(j,k),T_2}
\]
where $B(i, j)$ is a
function which gives the type of bond between atoms $i$ and $j$.

Unlike the Coulomb matrix, the Connectivity count representation is independent of molecular size, making it suitable for large systems. Coulomb matrices and Connectivity count features were determined from the implementation in the MOLML libraries.~\cite{collins2018constant}

\subsection{Excited state Data}
We took a subset of 21K+ molecules from the  QM9 dataset \cite{ramakrishnan2014quantum}. Because we have limited computational resources for TDDFT calculations, we restricted the molecules in the subset to those having 26 or fewer atoms of carbon (C), oxygen (O), nitrogen (N), fluorine (F), hydrogen (H), and sulfur (S). For each molecule, we computed the six lowest-lying excited state properties, specifically excitation energies and oscillator strengths, with B3LYP/TDDFT/DEF2-TZVPP  (high-level method for reference) and INDO/s (low-level method). These calculations were performed with the Gaussian 16 software \cite{g16}.

\subsection{$\Delta_{ML}$ training and test}
For each molecule, we considered the six lowest excited states and their corresponding oscillator strengths. A model was developed for each property in each state, resulting in a total of 12 models -- one for the excitation energy in the $n$-th state ($n=1$ to $6$) and the other for the corresponding oscillator strength. Each of the 12 models were developed using the KRR ML method. As mentioned previously, we also present here, 12 models obtained via the RF ML approach. There were 350 Coulomb Matrix (CM) and 132 Connectivity count (Co) descriptors for each model. The dataset was split randomly into training/test sets in the ratio 80/20 and, usual, in order to obtain the best performance of each ML technique, hyperparameters were optimized. We used a standard model selection procedure (grid search algorithm), paired with five-fold cross-validation (CV) for the hyperparamters optimization.
In KRR, hyperparameters such as regularization parameter gamma ($\gamma$),  Kernel function type (e.g Laplacian, radial basis function (rbf)), and kernel parameter($\alpha$) were optimized. 
For RF, the hyperparameters optimized were:  number of trees (n-estimators), tree depth (max-depth), minimum samples (min-sample-split and min-samples-leaf) and feature subsets (max-features). The optimum hyperparameters and their corresponding optimized value/function are given in the Supplementary Information.  

\section{Results and Discussion}\label{sec:Results}
\begin{figure}[h!]
    \centering
    \includegraphics[width=\textwidth]{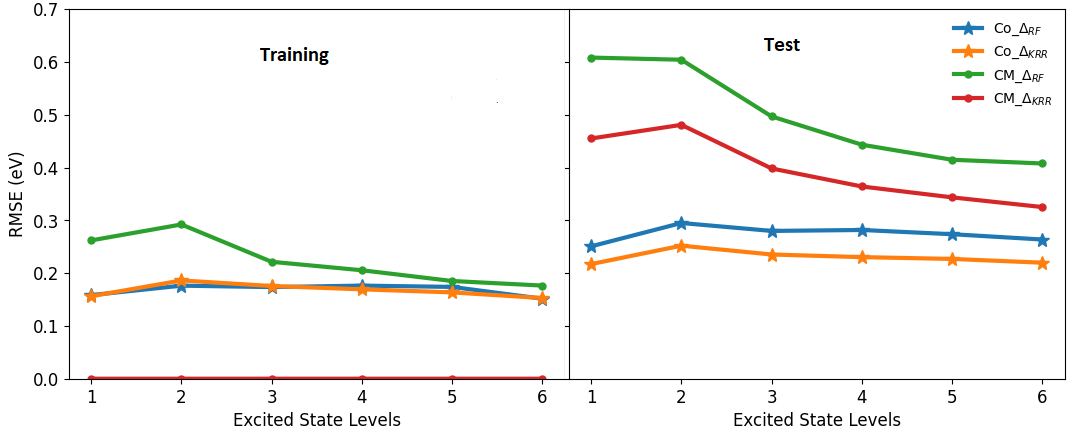}
    \caption{RMSE for training and test excitation energies datasets. The reference is the TDDFT excitation energies.}
    \label{del_ML1}
\end{figure}

In Figure \ref{del_ML1}, we present the Root-Mean-Square-Error (RMSE) of excitations energies for the training and test datasets using different descriptors ($CM$ and $Co$) in the KRR and RF models. It can be observed that KRR with $CM$ alone performed the best in training the excitation energy corrections for each state, achieving a training error which is essentially zero. With only the $Co$ descriptor, KRR and RF have similar errors: 0.17 eV. However, RF with the $CM$ descriptor alone performed the worst, with RMSE for the states ranging between 0.17 and 0.29 eV.
Next, the different models were tested using data containing more than 3,000 excitation energies (for each of the six states), which were not included in the training data. As shown in the right panel of Figure \ref{del_ML1}, KRR with $Co$ yielded the best predictions for the excitation corrections of each state, achieving an RMSE of less than 0.25~eV. In contrast, RF with $Co$ gave an RMSE in the range of 0.25 to 0.30~eV. However, the test data were not well reproduced using only the $CM$ features (with KRR or RF) as they have large prediction errors ranging from 0.34 to 0.60~eV. 

Overall, the models employing the $Co$ descriptor exhibited better transferability compared to those based on the Coulomb Matrix representation. The average difference in the RMSEs between the training and test data for the model with $Co$ was less than 0.1~eV whereas, for the model using the Coulomb matrix, the RMSE differences were greater than 0.2~eV.

\begin{figure}[h!]
    \centering
    \includegraphics[width=\textwidth]{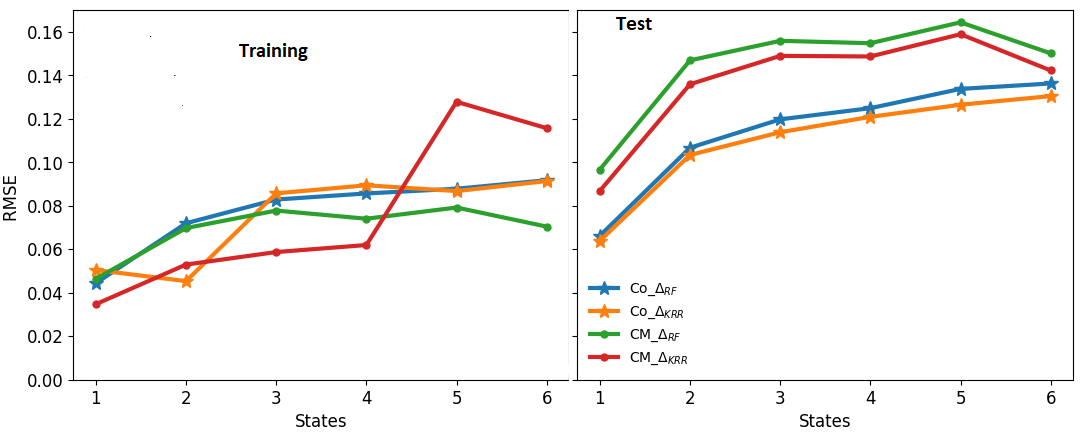}
    \caption{RMSE of transition dipole moments relative to the reference TDDFT results.}
    \label{del_ML2}
\end{figure}

Figure \ref{del_ML2} illustrates the results for the transition dipole moments for both the training and test datasets. For the former dataset, the ML corrections led to RMSEs of less than 0.14 across various models utilizing different featurizers. For the test dataset, models that used the $Co$ descriptor outperformed those that used $CM$. Nonetheless, all models demonstrated good transferability for predicting transition dipole moments, achieving predictions with an RMSE of 0.16 or less on the test set.
\subsection{Excitation energies and oscillator strength }
\begin{figure}
    \centering
    \includegraphics[width=\textwidth]{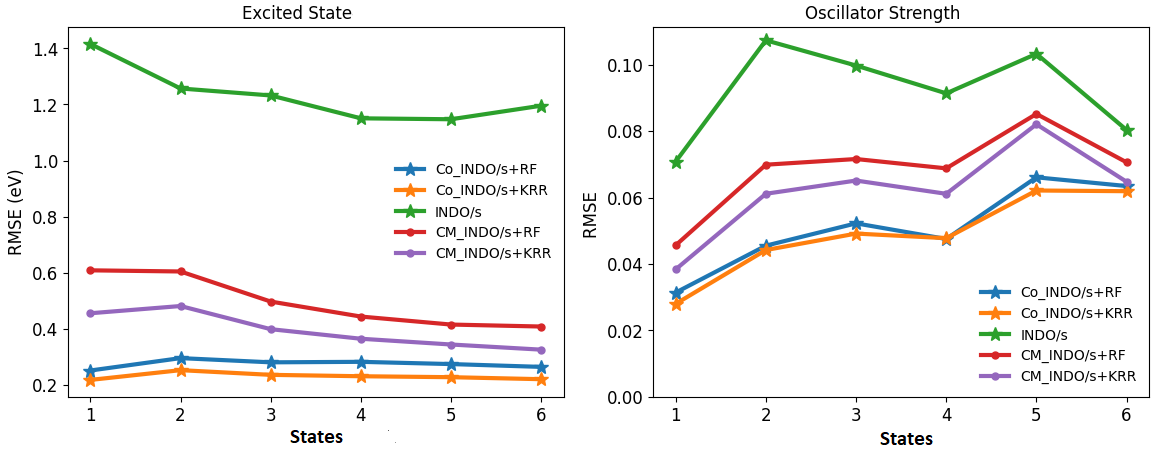}
    \caption{RMSE relative to TDDFT in (a) excitation energies and (b) oscillator strength of each state from the Orignial INDO/s and our new methods for the test sets.}
    \label{Ex}
\end{figure}

In this section, we present the predictions of excitation energies and oscillator strengths using our new methods (INDO/s + $\Delta_{\textsc{ML}}$). Figure \ref{Ex}a shows the RMSE of excitation energies for each state in the test sets. Our new methods demonstrate a significant improvement over the original INDO/s method. Specifically, the new method with the best performance, INDO/s+KRR with the $Co$ descriptor, exhibits an RMSE that is approximately 80\% lower than that of the original INDO/s method. The least-performing new method, INDO/s+RF with the $CM$ descriptor, shows a smaller reduction in error of about 50\% compared to the original RMSE of INDO/s.

Table \ref{tab1} summarizes the average prediction errors of excitation energies across all states from various methods relative to TDDFT. It reveals that KRR with the $Co$ descriptor achieves the lowest RMSE of 0.21 eV, while the errors from the other new methods range from 0.27 to 0.51 eV. In comparison, the original INDO/s method yields a larger RMSE of 1.28 eV.

The results and conclusions are similar for the oscillation strengths, but not as dramatic, see Fig.~\ref{Ex}b. Without the ML correction, oscillator strengths from INDO/s have RMSEs ranging from 0.07 to 0.11 with an average of 0.09 whereas, INDO/s+RF and INDO/s+KRR both have average RMSEs of 0.05 using the $Co$ descriptor, only a 44\% error reduction. The average RMSE error with the $CM$ descriptor is slightly larger (0.06 with KRR and 0.07 with RF) -- maximum error reduction of 33\%. 

\begin{figure}[!ht]
 
         \centering
         \includegraphics[width=\textwidth]{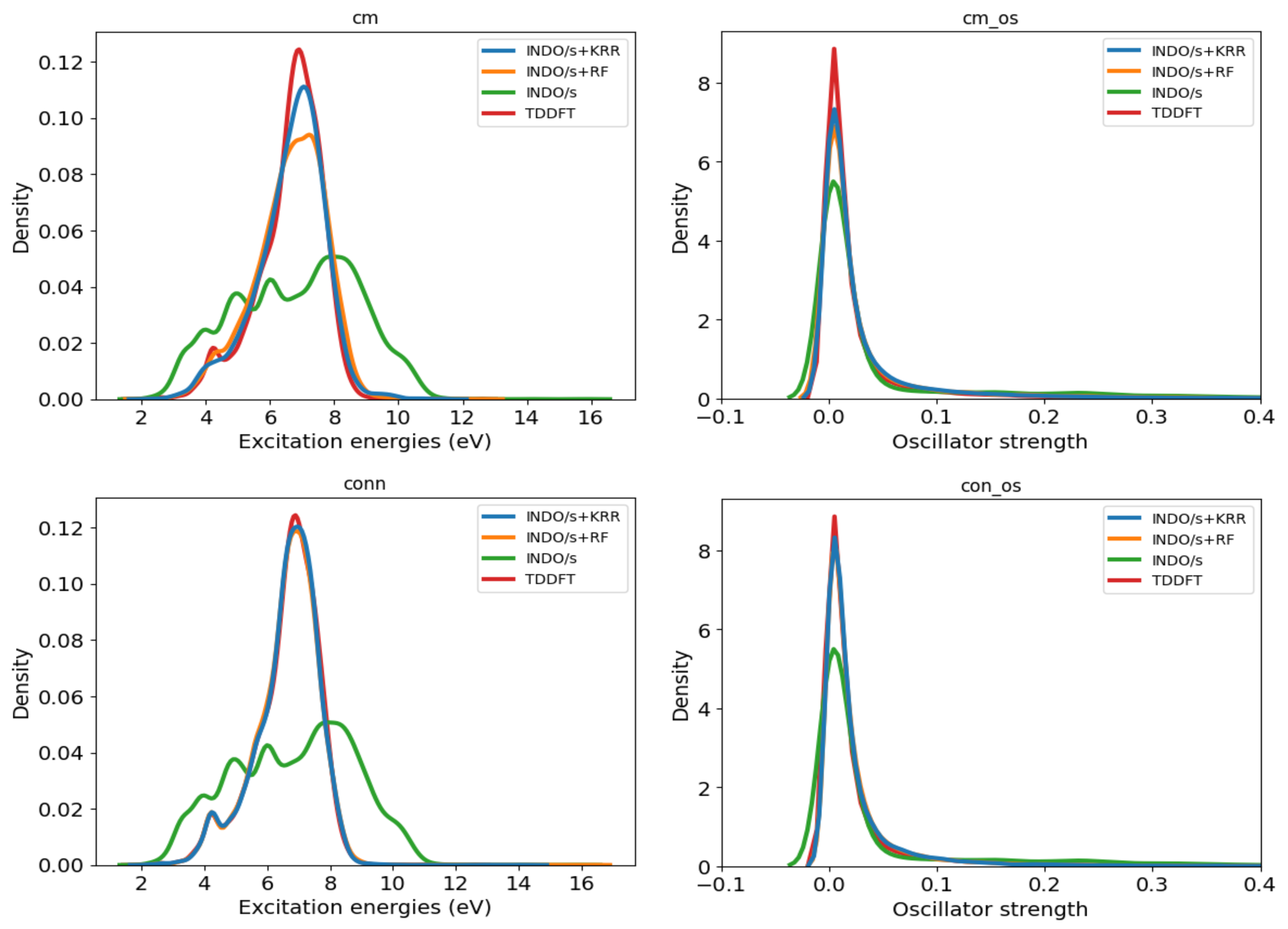}

        \caption{Kernel Distribution Estimation of: excitation energies with $CM$ descriptor (top left) with $Co$ descriptor (bottom left). Also, the distribution for the oscillation strengths with $CM$ (top right) and with $Co$ (bottom right) descriptors are shown for different methods. INDO/S+KRR matches TDDFT the best; it combines INDO/s with the $\Delta_{ML}$ from KRR.}
        \label{den}
\end{figure}

Additionally, we have assessed the accuracy of our new methods by plotting the density distribution of excitation energies for all states and molecules in the test set. Figure \ref{den} demonstrates that the spread of excitation energies and the peaks produced by models with the $Co$ descriptor closely match those from TDDFT. On the other hand, models utilizing the $CM$ exhibit minor deviations from the TDDFT results, with the highest peaks being lower than those from TDDFT. The original INDO/s method, however, shows significant deviation from the TDDFT reference, producing several peaks and broad bandwidth, in contrast to the only two peaks and narrow bands observed in the reference data. The oscillation strength distribution from KRR and RF, with the $Co$ descriptor, align closely with the TDDFT distribution. In contrast, the INDO/s oscillator strengths distribution deviates in both peak and spread compared to the TDDFT results.

\subsection{Absorption spectra}
\begin{figure}[h!]
 
         \centering
         \includegraphics[width=\textwidth]{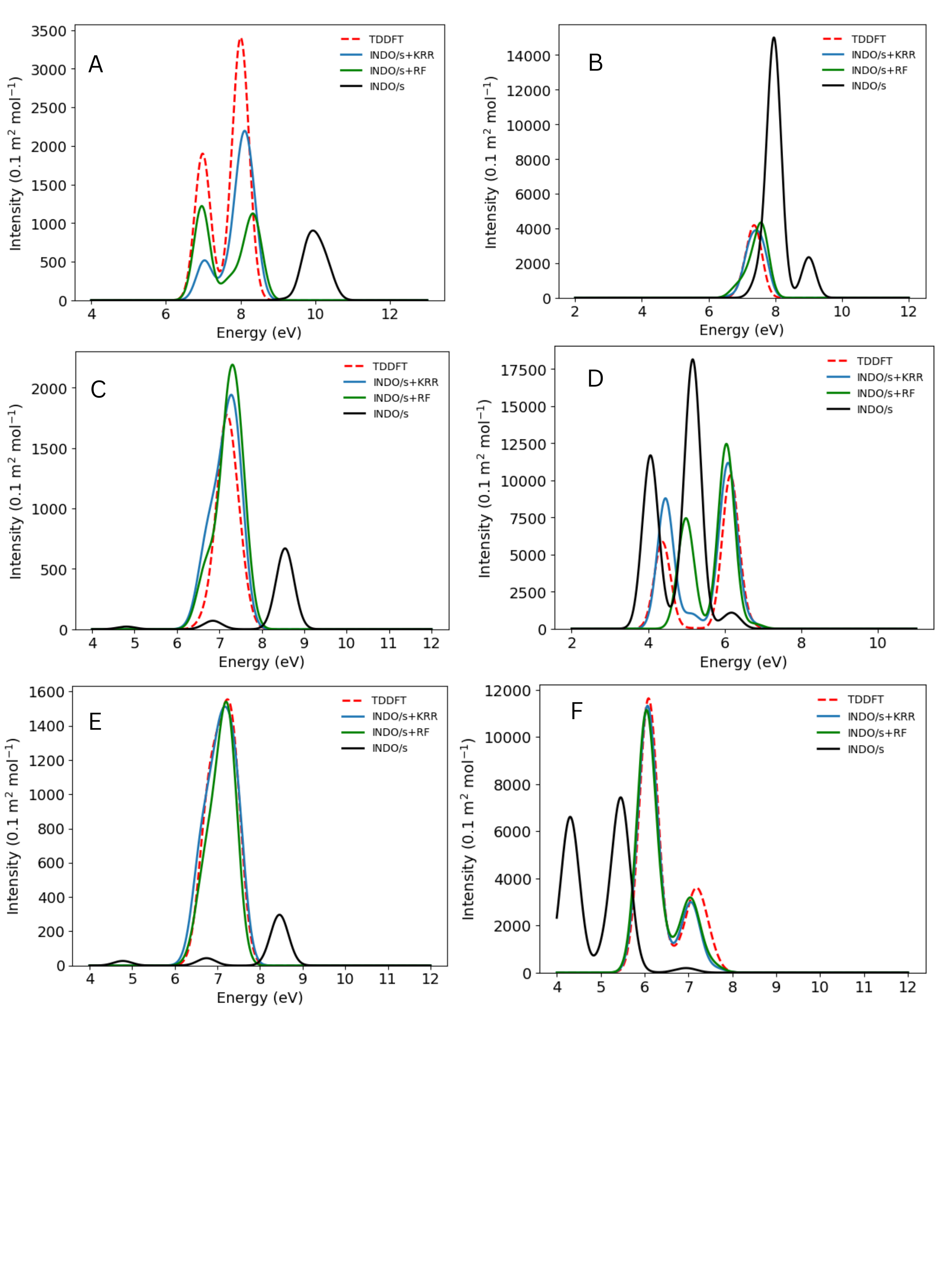}
        \caption{Comparing absorption spectra from our methods and original INDO/s to those from TDFFT of some molecules in the test set.}
        \label{spec}
\end{figure}

Figure \ref{spec} displays the absorption spectra computed using excitation energies and oscillator strengths for various test structures. We present spectra from new models (KRR and RF with the Co descriptor), as well as from the original INDO/s method and TDDFT, which serves as the reference method. In total, in this paper, we show UV-Vis spectra for six molecules; additional spectra and results from other models can be found in the supplementary document.

From the figures, it is evident that the new models predict spectra that closely align with the reference TDDFT spectra. The trends observed in the TDDFT spectra are reflected in the new models, with the peaks and bands showing a reasonable match. Figures B, C, D, and E demonstrate a consistent intensity match between the new models and TDDFT.

In contrast, the original INDO/s spectra deviate significantly from the reference. The patterns, intensities, and band ranges in the original INDO/s spectra do not align well with the reference. For example, in Figure 1, the INDO/s method produces a single low-intensity peak, whereas the reference reveals a high-intensity double-peaked spectrum. Additionally, in Figures A, B, C, and D , the highest peaks in the INDO/s spectra are blue-shifted compared to the highest peaks in the reference.

Overall, the new models represent a significant improvement over the original INDO/s method.

\section{Conclusion}
Machine learning-enhanced semi-empirical models, specifically INDO/s+KRR and INDO/s+RF, have been developed to provide near-time-dependent density functional theory (TDDFT) excitation energies, transition dipole moments, and UV-Vis absorption spectra for organic molecules at a low computational cost. Corrections were made to the properties of the original INDO/s method using machine learning models with various descriptor representations, resulting in the new methods INDO/s+KRR and INDO/s+RF. 

These new methods were evaluated using a test set of 3,600 molecules and demonstrated significantly better agreement with TDDFT results compared to the original INDO/s approach. Notably, the INDO/s+KRR model utilizing connectivity representation performed the best, showing good transferability across the test sets. It predicts TDDFT excitation energies and oscillator strengths with errors of approximately 0.2 eV and 0.05, respectively. In contrast, the original INDO/s method has corresponding errors of 1.1 eV and 0.08.  

\section{Acknowledgeement }
We thank CHPC South Africa and CINECA for granting access to a high-performance computer for calculations.
\clearpage
\newpage
\begin{table}
\centering
\caption{RMSE of excitation energies (in eV) and oscillation strengths for all states from various methods relative to TDDFT.}
\begin{tabular}{ l *{5}{d{4.5}} }
\toprule
& & \multicolumn{2}{c}{Training} & \multicolumn{2}{c}{Test}\\
\cmidrule(lr){3-4} \cmidrule(l){5-6}
& \mc{Model}& \mc{Connectivity} & \mc{CM} & \mc{Connectivity} & \mc{CM} \\
\midrule 
Excitation Energies & \mc{KRR} & 0.17 & 0.00 & 0.23& 0.40\\
       & \mc{RF} & 0.17 & 0.23 & 0.27 & 0.51 \\
       &\mc{$1.23^a$} & & & & \\
\addlinespace
\addlinespace
\addlinespace
Oscillator strength & \mc{KRR} & 0.03 & 0.04 & 0.05& 0.06\\
       & \mc{RF} & 0.03 & 0.03 & 0.05 & 0.07 \\
       & \mc{$0.09^b$} & & & &\\

\bottomrule
$^a$\footnotesize{RMSE for INDO/s excitation energies}\\
$^b$\footnotesize{RMSE for INDO/s oscillation strength}
\end{tabular} \label{tab1}
\end{table}
\bibliography{ezekiel.bib}



\end{document}